# UNIVERSAL BOUNDS ON THE SELFAVERAGING OF RANDOM DIFFRACTION MEASURES

Christof Külske[*]

Weierstraß-Institut für Angewandte Analysis und Stochastik
Mohrenstraße 39, D-10117 Berlin, Germany

August 31, 2001

ABSTRACT. We consider diffraction at random point scatterers on general discrete point sets in $\mathbb{R}^\nu$, restricted to a finite volume. We allow for random amplitudes and random dislocations of the scatterers. We investigate the speed of convergence of the random scattering measures applied to an observable towards its mean, when the finite volume tends to infinity. We give an explicit universal large deviation upper bound that is exponential in the number of scatterers. The rate is given in terms of a universal function that depends on the point set only through the minimal distance between points, and on the observable only through a suitable Sobolev-norm. Our proof uses a cluster expansion and also provides a central limit theorem.

## 1. INTRODUCTION AND SETUP

The study of the diffraction theory of 'ordered point sets' is a classical subject to physicists: Crystals produce sharp diffraction images, with bright spots known as Bragg peaks. It is known since the eighties (and mathematically well-understood by now) that this is also true for quasi-crystals ([Hof95b]). They posses long-range order but no translation symmetry and show geometrically intriguing diffraction patterns. Of course, sharp behavior of the scattering image occurs only in the limit of an infinite system of scatterers, much in analogy to the sharpness of phase transitions in statistical mechanics.

[*] Work supported by the DFG
Contact: kuelske@wias-berlin.de, http://www.wias-berlin.de/private/kuelske

What changes, if one is adding randomness ('disorder') to the picture? It is reasonable to expect that there should be a well-defined limit of the diffraction image when the number of scatterers tends to infinity, under natural assumptions. What assumptions exactly do we need mathematically? Do sample fluctuations matter? Do we have control over corrections to the infinite volume behavior when the number of scatterers is finite? There have been few mathematical papers about the first two questions (see however [BaaMoo98], [BaaHoe00], [Hof95a]), and no results at all about the finite volume behavior.

Suppose at first one is choosing the scatterers according to some translation-ergodic distribution while keeping the position fixed on a perfect crystal. Then the scattering images will converge to their disorder-averages in a distributional sense, by soft ergodicity arguments. This is true for almost any realization of the scatterers, using the ergodic theorem. Note however that these arguments do not provide any control over the finite volume corrections of the observed scattering image.

When one gives up translation invariance of the underlying structure or the distribution, there are not very many mathematical results in the literature. Ergodicity arguments are not available any more and one must resort to explicit methods. Subjecting the sites of a quasicrystal to an i.i.d. thermal motion leads to an infinite volume picture that is well-known from crystals: The intensity of the sharp peaks is reduced by a Debye-Waller factor with a diffuse background appearing. It is however not difficult to rigorously justify this kind of law-of-large number result when no control over the speed of convergence is required (see [Hof95a]). For mathematical results about the scattering at random tilings in the infinite volume limit we refer the reader to the recent review article of [BaaHoe00]. This paper provides a number of interesting and pedagogical examples and can also serve as a good introduction to mathematical scattering theory.

In the present paper we provide a contribution to the diffraction theory of random scatterers on general point sets by answering the question

**How large is the probability for a deviation of the scattering image of a finite portion of scatterers from its sample average?**

This is the probabilistically natural question for self-averaging and the experimentalist's question 'Is my system large enough?'. Our emphasis in this paper is that we do not assume a lattice structure, quasiperiodic structure or any symmetry of the set or of the distribution. For our results we only assume a minimal distance between the points of the reference point set. In particular all of our results hold for lattices or quasi-crystals. The answer then immediately leads to results about convergence of the scattering images for almost every realization, when it is combined with information about the behavior of the mean (without unnecessary assumptions as found in the literature).

We consider two types of randomness: A) We choose the scattering amplitudes according to a random distribution while keeping their locations fixed or B) subject them to thermal motion around their sites. We can treat both types in a unified way. In order to investigate the selfaveraging properties in a physically meaningful way, we adopt the following point of view. Fix an observable, modelling the counter used in an experiment, and look at the result of the corresponding measurement. Then estimate the probability distribution of the resulting quantity (that is random w.r.t sample fluctuations) in any finite volume. This provides much more information than the mere statement of the convergence of the scattering measures in the sense of distributions, for almost every realization of the scatterers, as it is conventionally done ([BaaMoo98], [BaaHoe00], [Hof95a]).

We note that, for a typical observable (e.g. Gaussian test function) the resulting expression will involve *all* autocorrelation coefficients of the array of scatterers and therefore does not trivially decompose into independent parts, even for scatterers that behave independently. In the language of statistical mechanics, *the observable produces an interacting system!* Treating the autocorrelation coefficients as individual random variables without using their dependence would suffice for a mere convergence result, but would lead to very bad large-deviation estimates. To deal with this interaction it will therefore be appropriate to employ (high-temperature) expansion methods from statistical mechanics, as will become clear soon. This shows the usefulness of such methods to give sharp explicit results, even in situations that do not a priori smell like dependent spin systems and Gibbs-measures. So, it would be nice if the paper could also serve as a motivation for probabilists and mathematical physicists who are sceptical about the use of expansions to take a closer look. In fact we restrict ourselves to the situation of independent scatterers to keep the technicalities down. To generalize the method to the case of weakly coupled scatterers is possible, but it would complicate the theorems, and make the general idea less transparent. We therefore leave it to a future paper.

In this setup we will provide general upper bounds on the probability that the measurement in finite volume deviates from its mean, and even provide explicit numerical values of the constants appearing. These estimates are universal in the sense that they depend only: 1) on the minimal distance between sites, but *not* on details of the point set; and 2) on the concentration of the observable, measured in a suitable Sobolev-norm of its Fourier-transform, but *not* on any more details. The fact that the estimate depends on the point set only through the minimal distance is important because one might want to be able to interpret the diffraction images without knowing beforehand the geometrical structure of the point set, while having some physical a priori-estimate on the mimimal distance.

Computing the average of a scattering image is simple and seeing whether it converges or not reduces to the knowledge of the autocorrelation structure already needed to understand the deterministic image of the point-set (see Appendix A).

Let us now define the models and state our results more precisely. We discuss the scattering image at infinity that is created by single-scattering at (a finite collection of) the point-scatterers described by the following random measures.

**Model A: Disordered scattering amplitudes, fixed positions**

We look at the complex random measure (*'random Dirac comb'*) given by

$$\rho_\Gamma(\eta) = \sum_{x \in \Gamma} \eta_x \delta_x \tag{1.1}$$

where $\delta_x$ denotes the Dirac-measure at the site $x$. The point set $\Gamma \subset \mathbb{R}^\nu$ is assumed to be countable. Here $\eta_x$ are random, possibly complex scattering amplitudes, independent over the sites $x \in \Gamma$. They are assumed to be bounded.

**Model B: Thermally dislocated scatterers**

Here we look at the scattering image of the random measure given by

$$\rho_\Gamma(\omega) = \sum_{x \in \Gamma} \delta_{x+\omega_x} \tag{1.2}$$

where $\omega_x$ are random ('thermal') dislocations taking values in $\mathbb{R}^\nu$, independent over the sites $x \in \Gamma$. They are assumed to be bounded, too.

Fix any finite volume $\Gamma_r \subset \Gamma$. Then, the object that contains all information about the scattering image of the points in $\Gamma_r$ is the *finite volume scattering measure* which is the Fourier-transform of the corresponding *finite volume autocorrelation measure*. (For a summary of the basic notions of mathematical scattering theory, see Appendix A and, for more details, e.g. Chapter II of [BaaHoe00].) Here, for Model A the autocorrelation measure in the finite volume $\Gamma_r$ is given by

$$\gamma_r^\eta := \frac{1}{|\Gamma_r|} \sum_{y \in \Gamma_r - \Gamma_r} \delta_y \sum_{\substack{x \in \Gamma_r: \\ x-y \in \Gamma_r}} \eta_x \eta_{x-y}^* \tag{1.3}$$

where the star denotes complex conjugate and the $y$-sum is over all difference vectors in $\Gamma_r$. Since we allow $\Gamma_r$ to be *any* finite set, we have chosen the natural normalization by the number of points (in contrast to [BaaHoe00]). This leads to simpler formulas in our theorems. For Model B we put

$$\gamma_r^\omega := \frac{1}{|\Gamma_r|} \sum_{x,x' \in \Gamma_r} \delta_{x-x'+\omega_x-\omega_{x'}} \tag{1.4}$$

for the finite volume autocorrelation measure. Suppose now a measurement on the scattered intensity is performed that is described by an observable $\varphi(k)$ in Fourier-space, modelling the counter. Usually it is assumed to be a real Schwartz function. The corresponding result of the measurement is then given by $\hat{\gamma}_r^\eta(\varphi) \equiv \int \hat{\gamma}_r^\eta(k) \varphi(k) dk$. Here the Fourier-transform of a tempered distribution $\gamma$ is defined by duality, $\hat{\gamma}(\varphi) = \gamma(\hat{\varphi})$, where $\hat{\varphi}$ denotes the Fourier-integral of the Schwartz-function $\varphi$ over $\mathbb{R}^\nu$. (For a quick reminder of the explanation for this and some comments, see Appendix A. For more expository details, see [BaaHoe00].) The *sample average* of the measurement is $\int \mu(d\eta)\hat{\gamma}_r^\eta(\varphi)$. This object has the correct normalization (by the total number of scatterers) to be able to converge to a well-defined limit, as it does of course for non-random scatterers on a lattice. (We remark that this average won't converge [e.g. for i.i.d. scatterers] along *any* sequence of volumes $\Gamma_r$, but e.g. an increasing sequence of balls or cubes works for crystals and quasi-crystals.)

**Main result for Model A**

Let us formulate the bound in the simplest form, which is suitable for the computation of explicit numbers bounding the probability of a large deviation. It makes explicit the uniformity of the large deviation upper-bound, independently of the set $\Gamma$ (other than the minimal distance between points in $\Gamma$), the form of the distribution (other than through uniform bounds on the magnitude of the scatterers), and the observable $\varphi$ (other than through a Sobolev-norm). We also provide a different version of the large deviation estimate in Chapter 4 under the name *'Addition to Theorem 1'*. It is slightly sharper in certain cases but less useful for direct application. In Chapter 4 we also give a corresponding *Central Limit Theorem*.

Now, to state the theorem we define the following Sobolev-norm involving integrals of derivatives up to the order of the dimension $\nu$, where we also introduce a scaling factor $a/2$. For a function $g : \mathbb{R}^\nu \to \mathbb{C}$ we put

$$\|g\|_{\nu,a} := \frac{1}{|B_1|} \sum_{k=0}^{\nu} \frac{1}{k!} \frac{1}{(a/2)^{\nu-k}} \int_{\mathbb{R}^\nu} \|d^k g(y)\| dy \tag{1.5}$$

Here $|B_1|$ denotes the volume of the $\nu$-dimensional unit ball. The symbol $d^k g(y) : (\mathbb{R}^\nu)^k \to \mathbb{R}^\nu$ denotes the $k$-th differential of $g$ at the point $y$ and $\|d^k g(y)\| := \sup_{|v_1|=\ldots|v_k|=1} |d^k g(y)[v_1, \ldots, v_k]|$ is the usual norm of a $k$-multilinear mapping, at any fixed point $y$, where $|v|$ denotes the Euclidean norm.

Then we have

**Theorem 1.** *Suppose that $\Gamma_r \subset \mathbb{R}^\nu$ is any finite set and denote the minimal distance between its points by $a$. Assume that $\eta = (\eta_x)_{x \in \Gamma}$ are (possibly complex) random variables, independent, but not necessarily identically distributed. Denote their distribution by $\mu$. Suppose the uniform bounds $|\mu(\eta_x)| \leq M < \infty$ and $|\eta_x - \mu(\eta_x)| \leq B < \infty$, for all $x \in \Gamma_r$, for $\mu$-a.e. realization.*

*Then the corresponding random scattering image $\hat{\gamma}_r^\eta(\varphi)$ in the finite volume $\Gamma_r$ obeys the universal large deviation estimate*

$$\mu\left(\left|\hat{\gamma}_r^\eta(\varphi) - \int \mu(d\eta)\hat{\gamma}_r^\eta(\varphi)\right| \geq \varepsilon\right) \leq 2\exp\left(-|\Gamma_r| \times J\left(\frac{\varepsilon}{K\|\hat{\varphi}\|_{\nu,a}}\right)\right) \tag{1.6}$$

*for any $\varepsilon > 0$, for any function $\varphi : \mathbb{R}^\nu \mapsto \mathbb{R}$, s.t. its Fourier-transform has finite norm $\|\hat{\varphi}\|_{\nu,a}$.*

*Here $K = 2MB + B^2$, and $J$ is a nonnegative, convex, strictly monotone function that is independent of the form of the distribution $\mu$ and the set $\Gamma$.*

*Remark.* The universal function $J : [0, \infty) \to [0, \infty)$ in the theorem has the form

$$J(\bar\varepsilon) = \begin{cases} \frac{16}{27D^2}\left(\left(1 + \frac{3}{4}D\bar\varepsilon\right)^{\frac{3}{2}} - 1 - \frac{9}{8}D\bar\varepsilon\right), & \text{if } \varepsilon \leq d(4 + 3Dd) \\ d\left(\bar\varepsilon - d\left(2 + dD\right)\right), & \text{else} \end{cases} \quad (1.7)$$

where the numerical constants can be chosen like $d = 0.0525$, and $D = 4.54 \cdot 10^3$. Note the asymptotics $J(\bar\varepsilon) \sim \frac{\bar\varepsilon^2}{8}$ for $\bar\varepsilon \downarrow 0$ and $J(\bar\varepsilon) \sim d\bar\varepsilon$ for $\bar\varepsilon \uparrow \infty$.

A structural and practical virtue of the form of the large-deviation estimate of Theorem 1 lies in the fact that its dependence on the observable $\varphi$ is formulated entirely in terms of the continuum-object $\|\hat\varphi\|_{\nu,a}$. All details of the set $\Gamma_r$ have disappeared! This Sobolev-norm can be computed (at least numerically) with little effort, and so one may easily derive explicit numbers.

*Remark.* Note the natural fact that the bound is scale-invariant in the following way: Suppose the counter is modelled by a probability density $\varphi_\sigma(k) = \sigma^{-\nu}\varphi_1(k/\sigma)$ in Fourier-space with variance ('precision of measurement') $\sigma^2$. (Think e.g. of a Gaussian!) Then, by scaling we have $\|\hat\varphi_\sigma\|_{\nu,a} = \|\hat\varphi_1\|_{\nu,a\sigma}$. So we have $\|\hat\varphi_\sigma\|_{\nu,a} \sim (a\sigma)^{-\nu} \int_{\mathbb{R}^\nu} |\hat\varphi_1(y)| dy$ with $\sigma \downarrow 0$, when $a$ is fixed (under the condition that the higher derivatives are integrable). This immediately controls the deterioration of our large deviation estimate when we make $\sigma$ smaller to increase the precision of measurement of the scattering image. (Without loss we could have chosen our length-scale in such a way that $a = 1$ from the beginning, so that the general statement is regained by rescaling the observable in $k$-space. We believe however that the present form of the theorem is more intuitive.)

*Remark.* The norm appearing is finite in particular for the commonly used Schwartz-test-functions. So our result in particular implies convergence-statements in the sense of tempered distributions. Suppose we are given an increasing sequence of volumes $\Gamma_r$. Then we immediately obtain the *strong law of large numbers* as a consequence of Theorem 1, saying that the centered autocorrelation measure applied to a test function $\varphi$ whose Fourier transform has finite norm converges to zero, for $\mathbb{P}$-a.e. $\eta$. This follows trivially by summing the exponential bound (1.6) over the volumes using the Borel-Cantelli Lemma.

*Remark.* The fact that the dependence on the observable $\varphi$, on the point-set $\Gamma$ and on the distribution $\mu$ can be expressed in terms of the handy quantity $K\|\hat\varphi\|_{\nu,a}$ is *not* a priori obvious. The occurrence of the norm however is not difficult to understand. It can be motivated by noting that $4(K\|\hat\varphi\|_{\nu,a})^2/|\Gamma_r|$ is an upper bound for the $\mu$-expectation of the square of the modulus inside the probability on the l.h.s. of (1.6). (This is seen using the independence of the scatterers by Fourier-transform, and substituting the norm-estimate of Proposition 3.) Believing in the corresponding Gaussian behavior, the small $\bar\varepsilon$-behavior given in (1.7) should follow. An essential part of the real proof consists in estimating *all the higher moments* contained in the Laplace-transform in terms of powers of $K\|\hat{\varphi}\|_{\nu,a}$. One can not expect a large deviation principle [which would in particular mean the existence of the limit $-\lim_r \frac{1}{|\Gamma_r|} \log \mu (|\ldots| \geq \varepsilon)$] without any assumptions on the set $\Gamma$ other than minimal distance. In fact, without further assumptions on $\Gamma$, the Laplace-transform won't converge.

## Main result for Model B

For the Model B of thermal dislocations the result is quite analogous. Here, however the Sobolev-norm of the *variation* of the Fourier-transform of the observable appears. Again, there will be a sharper version of this result in Chapter 4 that is called *'Addition to Theorem 2'*, and the *Central Limit Theorem*.

**Theorem 2.** *Suppose again that $\Gamma_r \subset \mathbb{R}^\nu$ is any finite set and denote the minimal distance between its points by $a$.*

*Suppose that the dislocations $\omega = (\omega_x)_{x \in \Gamma_r}$ have independent, not necessarily identical distribution $\mu$, such that $|\omega_x| \leq \delta < a/4$, for all $x \in \Gamma_r$, $\mu$-a.s.*

*Then the finite volume scattering image $\hat{\gamma}_r^\omega(\varphi)$ obeys the universal large deviation estimate*

$$\mu\left(\left|\hat{\gamma}_r^\omega(\varphi) - \int \mu(d\omega)\hat{\gamma}_r^\omega(\varphi)\right| \geq \varepsilon\right) \leq 2\exp\left(-|\Gamma_r| \times \tilde{J}\left(\frac{\varepsilon}{4\delta \, \|d\hat{\varphi}\|_{\nu,a-4\delta}}\right)\right) \qquad (1.8)$$

*The function $\tilde{J}$ has the same form as the function $J$ from Theorem 1 (see (1.7)), but with the slightly better constant $\tilde{D} = 4.38 \cdot 10^3 (\leq D)$ instead of $D$, and the same constant $d$.*

Here the appearing norm has the obvious meaning obtained by extending the previous definition (1.5) that was given for functions to linear functionals. It equals $\|dg\|_{\nu,a} = \frac{1}{|B_1|} \sum_{k=0}^{\nu} \frac{1}{k!} \frac{1}{(a/2)^{\nu-k}} \int_{\mathbb{R}^\nu} \|d^{k+1}g(y)\| dy$.

*Remark.* The restriction $\delta < a/4$ is only for simplicity. The more general statement of the *Addition to Theorem 2* stays true for any finite $\delta$. The fact that the estimate involves the Sobolev-norm of the *derivative* of $\varphi$ rather than the Sobolev-norm of $\varphi$ itself is due to the fact that $\gamma_r^\omega(\alpha)$ is non-random when the function $\alpha$ is a constant.

*Remark.* Note again the scale-invariance of the estimate, where of course the spatial distance $\delta$ must be rescaled, too: As for Model A, take a rescaled observable $\varphi_\sigma(k) = \sigma^{-\nu}\varphi_1(k/\sigma)$ in Fourier-space. Then we have $\delta \, \|d\hat{\varphi}_\sigma\|_{\nu,a-4\delta} = \delta\sigma \, \|d\hat{\varphi}_1\|_{\nu,a\sigma-4\delta\sigma}$. So, assuming that the higher derivatives are integrable, the quantity appearing in the large deviation estimate behaves like

$\delta \, \|d\hat{\varphi}_\sigma\|_{\nu,a-4\delta} \sim \delta(a-4\delta)^{-\nu}\sigma^{-(\nu+1)} \int_{\mathbb{R}^\nu} \|d\hat{\varphi}_1(y)\| dy \qquad$ with $\sigma \downarrow 0$ and $a, \delta$ fixed.

We conclude this introduction with an outline of the rest of the paper along with some ideas of the proof. In Chapter 2 we derive bounds on the Laplace transform of the centered random scattering measures, applied to some observable $\varphi$. To do so we look at this quantity as a (possibly complex) Hamiltonian of a spin-system. Here the random variables modelling the scatterers (resp. their dislocations) play the role of spins. The Laplace-transform then becomes a partition function. We can treat it by high-temperature expansion methods from statistical mechanics, under the assumption that the interaction be small. The smallness of the interaction of the spin-system we need for the expansion will be guaranteed by smallness of the Fourier-transform $\hat\varphi$ in a suitable norm. From the point of view of the expansion it is natural to introduce *discrete $\Gamma$-dependent norms*, so that we can control the terms of the expansion with constants that are independent of the structure of $\Gamma$. The resulting bounds for the Laplace-transforms including the computation of numerical constants are provided in Proposition 1 for Model A. In Chapter 3 the work of Chapter 2 is adapted to treat Model B. In Chapter 4 we state the sharpened results of the *'Additions to Theorems 1 and 2'* and the *Central Limit Theorem*, along with their proofs. They follow immediately from the norm-estimates on the Laplace-transform.

In Appendix A we recall the basic notions of scattering theory for point scatterers. In Appendix B we give estimates on our discrete $\Gamma$-dependent norms in terms of Sobolev-norms that depend on $\Gamma$ only through the minimal distance. This input is needed to show the uniformity in $\Gamma$ and the nice bounds given in Theorems 1 and 2.

## 2. NORM BOUNDS ON MOMENT GENERATING FUNCTION

In this chapter we use an expansion to derive bounds on the Laplace transform of the random variable in question. We will look at this random variable as a Hamiltonian of a spin system. We will formulate the bounds obtained in this chapter in terms of a suitable discrete norm that is close to what is needed for the proof of convergence, and compute numerical constants. These constants are obtained employing the known Kotecky-Preiss estimates for abstract polymer models. The result of this is found in Proposition 1.

Now, let us use the short notation

$$X_r(\alpha) \equiv |\Gamma_r| \ (\gamma_r^\eta(\alpha) - \mu(\gamma_r^\eta(\alpha))) \tag{2.1}$$

for the *nonnormalized centered autocorrelation measure* applied to the function $\alpha: \mathbb{R}^\nu \to \mathbb{C}$. This is the random variable in question. To derive bounds on its Laplace-transform and control the terms higher than second order in $\alpha$ we need a suitable norm. It turns out that the appropriate norm is the discrete $l^1$-type norm

$$\|\alpha\|_\Gamma := \sup_{x\in\Gamma} \sum_{z\in\Gamma} |\alpha(x-z)| \tag{2.2}$$

Of course, for $\Gamma = a\mathbb{Z}^d$ this is just an $l^1$-norm, for general $\Gamma$ it is slightly more complicated. The result of this chapter is then the following.

**Proposition 1.** *Suppose that the scatterers $\eta_x$ have independent, not necessarily identical distribution $\mu$, such that $|\mu(\eta_x)| \leq M < \infty$ and $|\eta_x - \mu(\eta_x)| \leq B < \infty$ for all $x \in \Gamma$, $\mu$-a.s.*

*Then there are universal constants $d > 0$, and $D < \infty$, independent of the set $\Gamma$ and the distribution $\mu$ (for given $K$), such that, whenever $\alpha : \mathbb{R}^\nu \to \mathbb{C}$ is such that $|\alpha(x)| = |\alpha(-x)|$ for all $x \in \Gamma$, and $K\|\alpha\|_\Gamma \leq d$, we have the estimate*

$$\left| \log \mu \left( e^{X_r(\alpha)} \right) - \frac{1}{2} \mu \left( X_r(\alpha) \right)^2 \right| \leq |\Gamma_r| D \times (K \|\alpha\|_\Gamma)^3 \tag{2.3}$$

*with $K := 2MB + B^2$.*

*The values of the constants can be chosen like $d = 0.0525$, and $D = 4.54 \cdot 10^3$.*

*Remark.* Note that the quadratic term in $\alpha$ under the modulus may not have a limit with $r \uparrow \infty$, for general sets $\Gamma$, even in the i.i.d. case. Much less need the higher moments of $X_r(\alpha)$ possess a limit. The essential point is however that all higher order terms in $\alpha$ are estimated *uniformly* in the set $\Gamma$. This uniformity follows from the cluster expansion error bounds and some explicit work.

*Proof.* We interpret $X_r(\alpha)$ as the (negative) Hamiltonian of a spin-system with spin-variables $\eta_x$, $x \in \Gamma_r$ and open boundary conditions. It is then most intuitive from the point of view of statistical mechanics to write it in the form

$$X_r(\alpha) = \sum_{\substack{\{x,z\} \subset \Gamma_r \\ x \neq z}} U_{x,z} + \sum_{x \in \Gamma_r} V_x \tag{2.4}$$

with the single-site potential

$$V_x = \alpha(0) \left( |\eta_x|^2 - \mu\left(|\eta_x|^2\right) \right) \tag{2.5}$$

and the pair potential

$$U_{x,y} = \alpha(x-y)\left(\eta_x \eta_y^* - \mu(\eta_x \eta_y^*)\right) + \alpha(y-x)\left(\eta_y \eta_x^* - \mu(\eta_y \eta_x^*)\right) \tag{2.6}$$

for $x \neq y$. Note that when $\alpha$ is the Fourier-transform of a real function, the pair potential is real. In general we allow it to be complex. Note that the potentials $V$ and $U$ are linear in the function $\alpha$. Note that in general the interaction will act between all pairs scatterers $\eta_x$. We look at the logarithm of the Laplace transform of $X_r(\alpha)$ which becomes the partition function of the spin-system. Then we want to compute the partition function of corresponding spin model to quadratic order in the strength of the interaction and control the remainder term. If we restrict ourselves to sufficiently 'small' $\alpha$ this will allow us to perform a cluster expansion, as we will see. This corresponds to 'small inverse temperature'. Such an expansion is in principle well-known in statistical mechanics, but we have to be careful about the precise assumptions we need on $\alpha$ and keep track of the constants appearing. It turns out that all we need to for the control is the quantity $K\|\alpha\|_\Gamma$.

Let us start. We note that

$$\|V_x\|_\infty \leq |\alpha(0)|B^2, \quad \|U_{x,y}\|_\infty \leq 2K|\alpha(x-y)| \tag{2.8}$$

Then we put $\mu[V](\cdot) := \mu(\cdot\ e^{\sum_{x\in\Gamma_r} V_x})/\mu(e^{\sum_{x\in\Gamma_r} V_x})$ to separate the single-site contributions and write

$$\begin{aligned}
\mu\left(e^{X_r(\alpha)}\right)/\mu(e^{\sum_{x\in\Gamma_r} V_x}) &= \mu[V]\left(\prod_{\{x,z\}\subset\Gamma_r}\left(e^{U_{x,z}} - 1 + 1\right)\right) \\
&= \sum_{T\subset\mathcal{B}_r} \mu\left(\prod_{\{x,z\}\in T}\left(e^{U_{x,z}} - 1\right)\right)
\end{aligned} \tag{2.9}$$

The set $\mathcal{B}_r$ describes the set of edges on the complete graph with vertices $\Gamma_r$. We write $T = P_1 \cup \cdots \cup P_n$ for the unique decomposition into connected components and call the $P_i$'s *polymers*. A polymer $P$ is thus of the form $P = \{\{x_1,z_1\},\{x_2,z_2\},\ldots,\{x_k,z_k\}\}$ and will be considered as a connected graph. There is the obvious notion of pairwise compatibility: $P_1,P_2$ are compatible iff they don't have any sites in common. So we write the last expression as a sum over pairwise compatible families of polymers with polymer-activities that depend on $\alpha$.

$$\mu\left(e^{X_r(\alpha)}\right) = \prod_{x\in\Gamma_r}\mu\left(e^{V_x}\right) \times \sum_{(P_1,\ldots,P_n)_c}\prod_{i=1}^n \rho_{P_i}(\alpha) \tag{2.10}$$

Here the polymer activity of a polymer is given by

$$\rho_P \equiv \rho_P(\alpha) = \mu[V]\left(\prod_{\{x,z\}\in P}\left(e^{U_{x,z}} - 1\right)\right) \tag{2.11}$$

This is the general formulation of a polymer partition function in an abstract polymer model. We want to perform the corresponding cluster-expansion for the logarithm of it. This is nothing but the Taylor-expansion when the polymer-activities are treated as independent (complex) variables $\rho_P$.

After this is done, we expand the activities $\rho_P$ to quadratic order as functions of $\alpha$. Expanding the exponential in powers of $\alpha$ gives the following. Let us write $g_l(s) = \sum_{i=l}^\infty \frac{s^i}{i!}$ for the remainder term of the Taylor-series of the exponential and use that $|g_l(s)| \leq g_l(|s|)$.

For a general polymer we have the bound

$$|\rho_P(\alpha)| \leq \prod_{\{x,z\}\in P} g_1(u_{x,z}) \tag{2.12}$$

with the abbreviation

$$u_{x,z} := \|U_{x,z}\|_\infty \tag{2.13}$$

using the uniform bound on the scatterers.

This is the bound we use to prove convergence of the expansion. We apply (a slight extension of) Proposition A.1 that can be found in [K01]. It says

**Proposition A.1 of [K01].** *Suppose that $\sum_{(P_1,\ldots,P_n)_c} \prod_{i=1}^n \rho_{P_i}$ is a polymer partition function, where: 'Polymers' $P$ are graphs on a set $\Gamma_r$ having at least one edge. Two polymers are called compatible if they have disjoint vertex sets. The sum is over pairwise compatible families of polymers. Assume that the (possibly complex) activities $\rho_P$ satisfy the bounds*

$$|\rho_P| \leq e^{-\sum_{b\in P}\tau_b} \quad \text{where} \quad \lambda := \sup_{x\in\Gamma_r} \sum_{y\in\Gamma_r: y\neq x} e^{-\tau_{x,y}} \leq \lambda^* \approx 0.110909 \tag{2.14}$$

*for some function $\tau_b = \tau_{x,y} \geq 0$ on the set of edges on $\Gamma_r$, where the above b-sum is over all edges of the graph $P$.*

*Then, the cluster expansion converges, i.e. the Taylor-series of the logarithm of the partition function has the representation*

$$\log \sum_{(P_1,\ldots,P_n)_c} \prod_{i=1}^n \rho_{P_i} = \sum_{\mathcal{C}} \Phi_{\mathcal{C}} \tag{2.15}$$

*where the sum is over indecomposable subsets $\mathcal{C}\subset\mathcal{P}$. 'Indecomposable' means that there do not exist nonempty $\mathcal{C}_1$ and $\mathcal{C}_2$ s.t. the pairs $P_1$, $P_2$ are always compatible for $P_1 \in \mathcal{C}_1$, $P_2 \in \mathcal{C}_2$. The weight $\Phi_{\mathcal{C}} = \sum'_{I:I\in\mathbb{N}^\mathcal{P}} c_I \prod_{P\in\mathcal{P}} \rho_P^{I_P}$ is the sum over all monomials in the Taylor-expansion corresponding to multi-indices $I$ with $I_P \geq 1$ for all $P \in \mathcal{C}$ and $c_I$ is the corresponding combinatorial factor, depending only on the incompatibility relation.*

*Moreover, we have the decay-estimate of the form*

$$\sum_{\mathcal{C}:\mathcal{C}\,icp\,P} |\Phi_{\mathcal{C}}| \left(\frac{\lambda^*}{\lambda}\right)^{|\mathcal{C}|} \leq a^*|P|, \quad \text{where} \quad a^* \approx 0.633 \tag{2.16}$$

*for any fixed $P$. Here the sum is over all clusters incompatible with $P$, i.e. containing at least one polymer incompatible with $P$ and we have put $|\mathcal{C}| = \sum_{P\in\mathcal{C}} |P|$ where $|P|$ is the number of bonds of the polymer $P$.*

The proof is the same as that provided in [K01]. Only the result was formulated for a translation-invariant setting, and applied as a technical tool in a different situation. (It relies on the the general Kotecky-Preiss estimate [KP86]. A simpler proof of this kind of result is given in [BoZa00].)

We note that in our case $\lambda$ is estimated from above by

$$\lambda \leq \lambda(\alpha) := \sup_{x \in \Gamma} \sum_{y \in \Gamma: y \neq x} g_1(u_{x,y}) \leq g_1(u) \tag{2.17}$$

where we have put

$$u := \sup_{x \in \Gamma} \sum_{y \in \Gamma: y \neq x} u_{x,y} \tag{2.18}$$

The second inequality of (2.17) follows from the positivity of the Taylor coefficients of $g_1$. This estimate explains the occurence of the norm $\|\cdot\|_\Gamma$. Such an estimate will be used over and over below.

To compute the logarithm of the Laplace transform up to quadratic order in $\alpha$ we need only keep clusters with at most two bonds. We get from the general estimate on cluster sums provided by (2.16) the bound

$$\left| \log \mu\left(e^{X_r(\alpha)}\right) - \sum_{x \in \Gamma_r} \log \mu\left(e^{V_x}\right) - \sum_{P:|P|=1,2} \Phi_{\{P\}} - \sum_{\substack{\{P_1,P_2\}:|P_1|=|P_2|=1, P_1 \neq P_2 \\ X(P_1) \cap X(P_2) \neq \emptyset}} \Phi_{\{P_1,P_2\}} \right|$$
$$\leq a^* |\Gamma_r| \left(\frac{g_1(u)}{\lambda^*}\right)^3 \qquad \text{for } u \leq \log(1+\lambda^*) \tag{2.19}$$

The hard part of the Taylor-expansion is now done by the general estimate. It remains to do some less elegant but elementary work: We still need to expand the three sums appearing under the modulus on the l.h.s. up to quadratic order in $\alpha$, estimate the remainder terms and verify that they can be estimated in terms of the norms we have introduced. The quadratic order term obviously produces $\frac{1}{2}\mu\left(X_r(\alpha)\right)^2$.

Now, the first sum is trivially estimated. Let us define the function $l(x) = -\log(1-x) - x = \sum_{k=2} x^k/k$. We have

$$\left|\log \mu\left(e^{V_x}\right) - \frac{1}{2}\mu\left(V_x^2\right)\right| = \left|\log(1+\mu(g_2(V_x))) - \mu(g_2(V_x)) + \mu(g_3(V_x))\right|$$
$$\leq l(g_2(v)) + g_3(v) \tag{2.20}$$

with

$$v := \sup_{x \in \Gamma} \|V_x\|_\infty \tag{2.21}$$

Here we used that $\mu(V_x) = 0$.

Let us come to the cluster sums. The cluster weights are obtained by comparing Taylor-coefficients (or by the inclusion-exclusion formula). One always has for single polymer clusters appearing in the second sum under the modulus of (2.19) that $\Phi_{\{P\}} =$ $\log(1+\rho_P)$. For these clusters we will write $\Phi_{\{P\}} = (\log(1+\rho_P) - \rho_P) + \rho_P$. Using $\mu(U_{x,y}) = 0$ we see that the activity of the single bond polymer $P = \{x,y\}$ is in fact of quadratic order in $\alpha$. [This is better than the application of the bound (2.12) which holds for all polymers would show.] Indeed,

$$|\rho_P(\alpha)| = \left| \frac{\mu(U_{x,y}g_1(V_x+V_y))}{\mu(e^{V_x+V_y})} + \mu[V](g_2(U_{x,y})) \right| \\ \leq u_{x,y}g_1(2v)e^{2v} + g_2(u_{x,y}) \qquad (2.22)$$

Therefore, $\log(1+\rho_P) - \rho_P$ is of fourth order, for both $|P| = 1, 2$. Thus we need to expand $\rho_P$ up to second order and control the third order error terms for both $|P| = 1, 2$. Finally the $\Phi_{\{P_1,P_2\}}$-term is of forth order, too. To control its magnitude it is convenient to use again Proposition A.1 using the improved bound (2.22).

Now, let us give some more details on the estimation of the error terms. To estimate the difference between the cluster weights appearing under the second sum in (2.19) and the corresponding activities we use $|\Phi_{\{P\}} - \rho_P| \leq l(|\rho_P|)$ to get

$$\left| \sum_{P:|P|=1,2} (\Phi_{\{P\}} - \rho_P) \right| \leq \sum_{\substack{\{x,y\} \\ x \neq y}} l\left( u_{x,y}g_1(2v)e^{2v} + g_2(u_{x,y}) \right) \\ + \sum_{y \in \Gamma_r} \sum_{\substack{x,z \in \Gamma_r \\ x \neq y, z \neq y, x \neq z}} l\Big(g_1(u_{x,y})g_1(u_{y,z})\Big) \qquad (2.23)$$

Using the fact that all the functions appearing have positive Taylor coefficients we may estimate the r.h.s. by

$$|\Gamma_r| \left( \frac{1}{2} l\left( ug_1(2v)e^{2v} + g_2(u) \right) + l\Big(g_1(u)^2\Big) \right) \qquad (2.24)$$

Next we need the error terms for the quadratic approximation on the polymer weights. Keeping the second order terms and using similar arguments as before we get for the single-bond polymer

$$\left| \rho_P(\alpha) - \mu(U_{x,y}(V_x+V_y)) - \frac{1}{2}\mu(U_{x,y}^2) \right| \\ \leq u_{x,y}\left[ 2vg_1(2v) + g_2(2v)e^{2v} \right] + \frac{1}{2}u_{x,y}^2 g_1(2v)(1+e^{2v}) + g_3(u_{x,y}) \qquad (2.25)$$

For a double-bond polymer $P = \{\{x,y\},\{y,z\}\}$ we get in a similar fashion

$$|\rho_P(\alpha) - \mu(U_{x,y}U_{y,z})| \leq u_{x,y}g_2(u_{y,z}) + u_{y,z}g_2(u_{x,y}) + g_2(u_{x,y})g_2(u_{y,z}) \\ + (u_{x,y} + u_{y,z})\left[ 2vg_1(2v) + g_2(2v)e^{2v} \right] + \frac{1}{2}u_{x,y}u_{y,z}g_1(3v)(1+e^{3v}) \qquad (2.26)$$

Summing over the polymer and using the positivity of the Taylor coefficients of $l, g_1, g_2$ we obtain

$$\begin{aligned}
\Bigg| \sum_{x \in \Gamma_r} \log \mu \left(e^{V_x}\right) &+ \sum_{P:|P|=1,2} \rho_{\{P\}} - \frac{1}{2} \mu \left(X_r(\alpha)^2\right) \Bigg| \leq |\Gamma_r| \Bigg( l(g_2(v)) + g_3(v) \\
&+ \frac{1}{2} u \left[2v g_1(2v) + g_2(2v) e^{2v}\right] + \frac{1}{4} u^2 g_1(2v)(1 + e^{2v}) + \frac{1}{2} g_3(u) \\
&+ 2u g_2(u) + g_2(u)^2 + 2u \left[2v g_1(2v) + g_2(2v) e^{2v}\right] + \frac{1}{2} u^2 g_1(3v)(1 + e^{3v}) \Bigg)
\end{aligned} \qquad (2.27)$$

Let us finally treat the last cluster sum under the modulus on the l.h.s. of (2.19) involving two single-bond polymers. For a pair of incompatible polymers $P_1, P_2$ one always has by the inclusion-exclusion formula that $\Phi_{\{P_1,P_2\}} = \log(1 + \rho_{P_1} + \rho_{P_2}) - \log(1 + \rho_{P_1}) - \log(1 + \rho_{P_2})$. The easiest way to treat this term here is by application of Proposition A.1 to the restricted polymer system that contains only single-bond polymers. We can use the improved second order bound (2.22). Denoting by $\Phi'_\mathcal{C}$ the corresponding cluster weights we thus have from (2.16) that

$$\sum_{\mathcal{C}: \mathcal{C} \text{ icp } P} |\Phi'_\mathcal{C}| \left(\frac{\lambda^*}{\lambda'}\right)^{|\mathcal{C}|} \leq a^* |P| \qquad (2.28)$$

with the same $a^* \approx 0.633$ and

$$\lambda' := \sup_{x \in \Gamma} \sum_{y \in \Gamma: y \neq x} |\rho_{\{x,y\}}| \leq u g_1(2v) e^{2v} + g_2(u) \qquad (2.29)$$

So we get

$$\sum_{\substack{\{P_1,P_2\}: |P_1|=|P_2|=1, P_1 \neq P_2 \\ X(P_1) \cap X(P_2) \neq \emptyset}} |\Phi_{\{P_1,P_2\}}| \leq \frac{a^*}{(\lambda^*)^2} |\Gamma_r| \left(u g_1(2v) e^{2v} + g_2(u)\right)^2 \qquad (2.30)$$

Collecting terms we arrive at the final estimate

$$\left| \log \mu \left(e^{X_r(\alpha)}\right) - \frac{1}{2} \mu \left(X_r(\alpha)^2\right) \right| \leq |\Gamma_r| h(u,v) \qquad (2.31)$$

with

$$\begin{aligned}
h(u,v) = \frac{a^*}{(\lambda^*)^3} g_1(u)^3 &+ l(g_2(v)) + g_3(v) \\
&+ \frac{1}{2} u \left[2v g_1(2v) + g_2(2v) e^{2v}\right] + \frac{1}{4} u^2 g_1(2v)(1 + e^{2v}) + \frac{1}{2} g_3(u) \\
&+ 2u g_2(u) + g_2(u)^2 + 2u \left[2v g_1(2v) + g_2(2v) e^{2v}\right] + \frac{1}{2} u^2 g_1(3v)(1 + e^{3v}) \\
&+ \frac{1}{2} l \left(u g_1(2v) e^{2v} + g_2(u)\right) + l\left(g_1(u)^2\right) + \frac{a^*}{(\lambda^*)^2} \left(u g_1(2v) e^{2v} + g_2(u)\right)^2
\end{aligned} \qquad (2.32)$$

Now we use that $u \leq 2K\|\alpha\|_\Gamma$ and $v \leq K\|\alpha\|_\Gamma$. So, $K\|\alpha\|_\Gamma \leq \frac{1}{2}\log(1+\lambda^*) =: d \approx 0.05258$ implies that the cluster expansion is convergent. We get for these $K\|\alpha\|_\Gamma$ the third order norm-estimate on the higher terms in the form

$$h(u,v) \leq h\left(2K\|\alpha\|_\Gamma, K\|\alpha\|_\Gamma\right) \leq D(K\|\alpha\|_\Gamma)^3 \qquad (2.33)$$

with $D := \sup_{x:0\leq x\leq d} \frac{h(2x,x)}{x^3} = \frac{h(2d,d)}{d^3}$. This is clear by the positivity of the Taylor-coefficients of $h(x)$. It is then a trivial matter to compute the constant $D$ given in the claim of the proposition numerically. We get $D \leq, \approx 4352 + 63 + 124 \leq 4540$ where the first number gives a bound on the first term, the last number a bound on the last term, and the middle number a bound on the remaining terms of (2.32). This shows that the error term coming from the estimation of the higher order terms in the cluster expansion that depends on $u$ alone provides by far the main contribution.

*Remark.* Of course one cannot expect the series to converge without any smallness assumptions on $\|\alpha\|_\Gamma$. In fact, for $\Gamma = \mathbb{Z}^\nu$ with $\nu \geq 2$ and $\alpha(x) = J1_{|x|=1}$ we are back to the usual ferromagnetic nearest-neighbor Ising-model, and the series is known to diverge for large $J$ due to the existence of a phase transition.

## 3. THERMAL DISLOCATIONS

It is not too difficult to go through the proof given in the previous section to accommodate the case of Model B of thermal dislocations. There are some changes, however. First of all, we need a different norm estimating the variation of the (Fourier transform of) the observable w.r.t. variations up to the magnitude $\delta$. We define the semi-norm

$$\|\alpha\|_{\Gamma,\delta} := \sup_{x\in\Gamma} \sum_{\substack{y\in\Gamma \\ y\neq x}} \sup_{\substack{z,z'\in\mathbb{R}^\nu \\ |z|,|z'|\leq 2\delta}} \left|\alpha(x-y+z) - \alpha(x-y+z')\right| \qquad (3.1)$$

Note that this seminorm vanishes on constant functions. We denote

$$Y_r(\alpha) \equiv |\Gamma_r| \; (\gamma_r^\omega(\alpha) - \mu(\gamma_r^\omega(\alpha))) \qquad (3.2)$$

for the nonnormalized centered autocorrelation measure applied to the function $\alpha$.

Then we have a norm-estimate on the Laplace-transform that is analogous to Proposition 1. The result is the following

**Proposition 2.** *Suppose that the dislocations $\omega = (\omega_x)_{x\in\Gamma}$ have independent, not necessarily identical distribution $\mu$, such that $|\omega_x| \leq \delta < \infty$, for all $x \in \Gamma$, $\mu$-a.s.*

*Then there are universal constants $d > 0$, and $\tilde{D} < \infty$, independent of the set $\Gamma$ and the distribution $\mu$, such that, whenever $\alpha : \mathbb{R}^\nu \to \mathbb{C}$ is small enough such that $\|\alpha\|_{\Gamma,\delta} \leq d$ and $|\alpha(x)| = |\alpha(-x)|$ for all $x$, we have the estimate*

$$\left| \log \mu \left( e^{Y_r(\alpha)} \right) - \frac{1}{2} \mu \left( Y_r(\alpha) \right)^2 \right| \leq |\Gamma_r| \tilde{D} \, \|\alpha\|_{\Gamma,\delta}^3 \tag{3.3}$$

*The values of the constants can be chosen like $d = 0.0525$ (same as in Proposition 1), and $\tilde{D} = 4.38 \cdot 10^3 (\leq D)$.*

*Proof.* We follow the lines of the proof given in the previous section. Our 'Hamiltonian' now becomes

$$Y_r(\alpha) = \sum_{\substack{\{x,z\} \subset \Gamma_r \\ x \neq z}} U_{x,z} \tag{3.4}$$

with the pair potential

$$\begin{aligned} U_{x,y} =& \alpha\big(x - y + \omega_x - \omega_y\big) - \mu\Big(\alpha\big(x - y + \omega_x - \omega_y\big)\Big) \\ &+ \alpha\big(y - x + \omega_y - \omega_x\big) - \mu\Big(\alpha(y - x + \omega_y - \omega_x)\Big) \end{aligned} \tag{3.5}$$

for $x \neq y$. Note that there is no single site potential this time, since the corresponding expression vanishes for $x = y$. We note that

$$\|U_{x,y}\|_\infty \leq 2 \sup_{\substack{z,z' \in \mathbb{R}^\nu \\ |z|,|z'| \leq 2\delta}} \left| \alpha(x - y + z) - \alpha(x - y + z') \right| \tag{3.6}$$

So we have

$$u' := \sup_{x \in \Gamma} \sum_{y \in \Gamma: y \neq x} u_{x,y} \leq 2\|\alpha\|_{\Gamma,\delta} \tag{3.7}$$

Now the steps of the proof of Proposition 1 stay true, leading to formula (2.31) with $v = 0$.

$$\left| \log \mu \left( e^{Y_r(\alpha)} \right) - \frac{1}{2} \mu \left( Y_r(\alpha)^2 \right) \right| \leq |\Gamma_r| h(u', v = 0) \tag{3.8}$$

with the function $h$ given in (2.32). The constant $d$ stays the same and for the constant $\tilde{D}$ we get the better value $\tilde{D} = \frac{h(2d,0)}{d^3} \leq \approx 4352 + 10 + 12 \leq 4380$. This shows that we get essentially the same constant as that of Proposition 1 and the diagonal terms didn't do much harm.

## 4. MORE BOUNDS, CLT, AND FINAL PROOFS

In this chapter we state the more detailed versions of Theorems 1 and 2, along with their proofs and also provide the Central Limit Theorem. In particular the results will still contain the discrete $\Gamma$-dependent norms (2.2) resp. (3.1). We have preferred here to write the estimates in terms of the autocorrelation-measure applied to a function (rather than its Fourier-transform). We will use here the notations from Chapter 2 for Model A and Chapter 3 for Model B.

Let us start with the result for Model A. Suppose that $\alpha$ is a given function on $\mathbb{R}^\nu$ with $|\alpha(x)| = |\alpha(-x)|$ for all $x \in \mathbb{R}^\nu$. Recall the definition of the discrete norm $\|\alpha\|_\Gamma$ given in (2.2). Recall the notation $X_r(\alpha) \equiv |\Gamma_r| \, (\gamma_r^\eta(\alpha) - \mu \, (\gamma_r^\eta(\alpha)))$ for the nonnormalized centered autocorrelation measure applied to the function $\alpha$. In this situation the following result holds.

**Addition to Theorem 1.** *We have the large deviation estimate*

$$\mu\Big(|X_r(\alpha)| \geq \varepsilon |\Gamma_r|\Big) \leq 2 \exp\left(-|\Gamma_r| \times j_{d,D}\Big(\frac{\varepsilon}{K\|\alpha\|_\Gamma}; s_r\Big)\right) \tag{4.1}$$

*for all $\varepsilon > 0$, where $s_r = \frac{1}{|\Gamma_r|}\mu\left(X_r^2\left(\frac{\alpha}{K\|\alpha\|_\Gamma}\right)\right) \leq 4$.*

*Here, for fixed $s \geq 0$ the function $j_{d,D}(\,\cdot\,; s) : [0, \infty) \to [0, \infty)$ has the form*

$$j_{d,D}(\bar{\varepsilon}; s) = \begin{cases} \frac{1}{108D^2}\left((12D\bar{\varepsilon} + s^2)^{\frac{3}{2}} - 18D\bar{\varepsilon}s - s^3\right), & \text{if } \bar{\varepsilon} \leq d(s + 3Dd) \\ d\left(\bar{\varepsilon} - d\left(\frac{s}{2} + dD\right)\right), & \text{else} \end{cases} \tag{4.2}$$

*where $d > 0$, $D < \infty$ are the same numerical constants as in Theorem 1. It is convex, nonnegative, strictly increasing in $\bar{\varepsilon}$. It is decreasing in $s$ and in $D$, and increasing in $d$.*

*Remark.* The statement is stronger than the simpler one given in Theorem 1 in two ways. First of all, the Sobolev-norm $\|\cdot\|_{\nu,a}$ appearing therein is replaced by the sharper norm $\|\cdot\|_\Gamma$ introduced in (2.2). This is only very minor because the Sobolev-norm will be used in practical applications. Next, we have kept the normalized variance $s_r$. Usually $s_r$ will be of the order unity, e.g. when $\Gamma$ is a lattice and the scatterers are i.i.d. There can however be cases of sets $\Gamma_r$ and functions $\alpha$ for which this quantity will go to zero with $r \uparrow \infty$.

Now, putting together the two pieces of information $s_r \leq 4$ and $\|\cdot\|_\Gamma \leq \|\cdot\|_{\nu,a}$ (see Appendix B, Proposition 3) the simplified statement of Theorem 1 immediately follows, by the monotonicity of the function $j$ in $s$.

The situation is completely analogous for Model B of thermal dislocations. To formulate the corresponding statement we recall definition (3.2) for the non-normalized autocorrelation-measure applied to $\alpha$. Recall the discrete $\Gamma$-dependent semi-norm (3.1). Then our result reads as follows.

**Addition to Theorem 2.** *Suppose that $\delta < \infty$. Then we have*

$$\mu\Big(|Y_r(\alpha)| \geq \varepsilon|\Gamma_r|\Big) \leq 2\exp\Big(-|\Gamma_r| \times j_{d,\tilde{D}}\Big(\frac{\varepsilon}{\|\alpha\|_{\Gamma,\delta}}; q_r\Big)\Big) \tag{4.3}$$

*for all $\varepsilon > 0$, where $q_r = \frac{1}{|\Gamma_r|}\mu\Big(Y_r^2\Big(\frac{\alpha}{\|\alpha\|_{\Gamma,\delta}}\Big)\Big)$. Again $d > 0$, $\tilde{D} < \infty$ are the same numerical constants as in Theorem 2 and the function $j_{d,\tilde{D}}$ is given in (4.2).*

*Remark.* The statement of Theorem 2 follows from here by $q_r \leq 4$ and the norm estimate given in Appendix B, Proposition 4.

Looking at the variable on the central-limit scale we get the following result.

**Theorem 3.** *Suppose that $\lim_{r\uparrow\infty} \mu\big(X_r^2(\alpha)\big)|\Gamma_r|^{-\frac{2}{3}} = \infty$. Then the standardized variable $X_r(\alpha)\big(\mu\big(X_r^2(\alpha)\big)\big)^{-\frac{1}{2}}$ converges weakly to a standard Gaussian distribution. The same statement holds for $Y_r$ replacing $X_r$.*

Finally we give the proofs.

*Proof of the Addition to Theorem 1.* Assuming the uniform estimates on the Laplace transform provided in Proposition 1 it is a trivial matter to derive the large deviation upper bound. Indeed, by the exponential Chebychev inequality we have

$$\begin{aligned}\mu\left(\pm X_r(\alpha) \geq |\Gamma_r|\varepsilon\right) &\leq \inf_{t:0\leq t\leq d} e^{-\frac{\varepsilon}{K\|\alpha\|_\Gamma}|\Gamma_r|t}\mu\left(e^{\pm X_r(t\alpha/(K\|\alpha\|_\Gamma))}\right)\\ &\leq \exp\left(-|\Gamma_r| \times \sup_{t:0\leq t\leq d}\left(\frac{\varepsilon}{K\|\alpha\|_\Gamma}t - \frac{1}{|\Gamma_r|}\mu\left(X_r^2\left(\frac{\alpha}{K\|\alpha\|_\Gamma}\right)\right)\frac{t^2}{2} - Dt^3\right)\right)\end{aligned} \tag{4.4}$$

Call the function appearing in the exponent in the bound

$$j_{d,D}(\bar{\varepsilon}; s) := \sup_{t:0\leq t\leq d}\left(\bar{\varepsilon}t - \frac{st^2}{2} - Dt^3\right) \tag{4.5}$$

Observe that $j_{d/\lambda,\lambda^3 D}(\lambda\bar{\varepsilon}; \lambda^2 s) = j_{d,D}(\bar{\varepsilon}; s)$. A simple computation shows then that $j$ has in fact the explicit form (4.2) given in the Addition to the Theorem 1. (In the small $\varepsilon$-range the maximizer is $t = \frac{-s+\sqrt{12D\varepsilon+s^2}}{6D}$, in the large $\varepsilon$-range the maximizer is $t = d$.) $j$ is convex and nonnegative. The monotonicity properties claimed in the Addition to Theorem 1 are now immediate by formula (4.5). Finally, it is a simple exercise to see that $s_r \leq 4$, using the independence of the scatterers.

The proof of the Addition to Theorem 2 is the same.

*Proof of Theorem 3.* From the bound on the error for the quadratic approximation of the Laplace transform given in Proposition 1 follows immediately that, for all fixed $t \in \mathbb{C}$ we have that $\lim_{r \uparrow \infty} \log \mu \left( \exp \left( t X_r(\alpha) \left( \mu \left( X_r^2(\alpha) \right) \right)^{-\frac{1}{2}} \right) \right) = t^2/2$, under the assumptions of the theorem. This shows the claim. The proof for Model B is identical.

# APPENDIX A: SCATTERING THEORY FOR POINT SCATTERERS

Let us briefly recall the basic elementary formulae of scattering theory that describe the connection between the autocorrelation $\gamma_r^\eta$ (resp. $\gamma_r^\omega$ ) and the scattering image. For notational concreteness we only consider Model A (fixed locations). (For more on this see [Hof95b], [BaaHoe00]). Suppose a beam with wavelength $\lambda_b$ hits the finite collection of point-scatterers located in the finitely many points $\Gamma_r$. Denote by $e_0 \in \mathbb{R}^\nu$ the incoming direction (where $|e_0| = 1$ is a unit vector). The modulus of the scattering amplitude $\eta_x$ gives the amplitude of the scattered wave and the phase of $\eta_x$ gives a local phase shift at the site $x$. Consequently the intensity of radiation scattered elastically in the direction $e$ is given by $\left| \sum_{x \in \Gamma_r} \eta_x e^{ik \cdot x} \right|^2$ with $k = 2\pi(e - e_0)/\lambda_b$. To understand the l.h.s. of this formula take $\eta_x \equiv 1$ and note that in this case $k \cdot x$ is the phase difference of a beam scattered at site $x$ relative to that of a beam scattered at a (hypothetical) scatterer at site 0.

Multiplying the intensity by a test-function $\varphi(k)$ (that models the sensitivity of a counter) and normalizing by the number of scatterers then leads to the quantity $\gamma_r^\eta(\hat{\varphi}) = \hat{\gamma}_r^\eta(\varphi)$. Here we choose the convention $\hat{\varphi}(x) = \int_{\mathbb{R}^\nu} e^{ix \cdot k} \varphi(k) dx$ to define the Fourier-transform of a Schwartz-function $\varphi$. The Fourier-transform of a tempered distribution is then defined by duality.

So, when one is interested in the infinite volume limit, one likes to look ([Hof95b]) at the scattering measures $\hat{\gamma}_r^\eta$ in the sense of (tempered) distributions and is interested in the weak limit $r \uparrow \infty$, i.e. $\lim_{r \uparrow \infty} \hat{\gamma}_r^\eta(\varphi)$ where $\varphi$ is a Schwartz function. Then, if a limiting distribution exists at all, it can have a discrete part, an absolutely continuous and a singular continuous part, the discrete part (Bragg peaks) caused by 'order', the continuous parts showing diffuse scattering caused by 'disorder' of the scatterers. This is in analogy to statistical mechanics where sharp phase transitions occur only in the infinite volume.

## Disorder-averages of diffraction measures

Our theorems give us good control over the fluctuations of the scattering measures $\hat{\gamma}_r^\eta$. The estimates are independent of the behavior of the mean, and the nature of the limimiting distribution, if it exists. To compute their disorder averages of the scattering measures is a trivial matter. We get

**Model A:**  $\int \mu(d\eta)\hat\gamma_r^\eta(\varphi) = \hat\gamma_r^m(\varphi) + \frac{1}{|\Gamma_r|}\sum_{x\in\Gamma_r}\left(\mu(|\eta_x|^2) - |\mu(\eta_x)|^2\right) \times \hat\varphi(0)$

where $\mu(\eta_x) =: m_x$ is the mean-value of the scattering amplitude and $m = (m_x)_{x\in\Gamma}$. The first term describes the scattering image of a system where the scattering amplitudes have been replaced by their means and the second term a homogenous diffuse background. So we see, that a.s. convergence for the averaged scattering images holds if and only if the two individual terms converge. This is true for $\Gamma$ a crystal or quasicrystal and $\eta_x$ are i.i.d., with e.g. $\Gamma_r$ being increasing balls. The latter statement follows since a crystal or quasicrystal is known to possess a natural autocorrelation function. To see how to construct an example of independent but not identically distributed scatterers on a quasicrystal for which the mean converges, see Paragraph 7 of [BaaMoo98]. On the other hand, it is simple to construct examples of systems on lattices with prescribed convergence/non-convergence of each of the two terms along a given sequence of volumes. This is done by choosing the distribution of $\eta_x$'s in a non-homogenous way; think of a sparse sequence of increasing volumes $\Gamma_r$ and choose two different distributions in the annuli $\Gamma_{r+1}\backslash\Gamma_r$ for $r$ even resp. $r$ odd. Still, also in these examples without convergence of the mean, under the assumption of uniform boundedness of the distribution, selfaveraging in the sense of Theorem 1 would hold.

**Model B:** $\int \mu(d\omega)\hat\gamma_r^\omega(\varphi) = \frac{1}{|\Gamma_r|}\sum_{x\neq x'\in\Gamma_r}\mu\left(\hat\varphi(x - x' + \omega_x - \omega_{x'})\right) + \hat\varphi(0)$

Again, one can construct artificial distributions of dislocations of scatterers on a lattice such that this expression does not have a well-defined limit. Choose e.g. $\omega_x \equiv 0$ for $x$ in the annuli $\Gamma_{r+1}\backslash\Gamma_r$ for $r$ even, and a non-trivial bounded law for $\omega_x$ for $r$ odd. Still, self-averaging holds.

However, if the $\omega_x$'s are i.i.d. with single-site distribution $\mu$ we get from this $\int \mu(d\omega)\hat\gamma_r^\omega(k) = \hat\gamma_r^0(k)|\hat\mu(k)|^2 + \left(1 - |\hat\mu(k)|^2\right)$ where $\hat\gamma_r^0(k)$ is the density of the Fourier-transform of the autocorrelation with all the scatterers sitting at their sites in $\Gamma_r$ and $|\hat\mu(k)|^2 = |\mu(e^{i\omega_x\cdot k})|^2$ is the famous Debye-Waller factor reducing the intensity of the reflexes.

## APPENDIX B: NORM ESTIMATES

Finally we like to give the norm-estimates along with their proofs that are needed to obtain the final form of the theorems as they are stated in the introduction.

**Proposition 3.** *Suppose that $\Gamma\subset\mathbb{R}^\nu$ and the number $a$ is a bound on the minimal distance between the points in $\Gamma$. Then we have the bound on the discrete $\Gamma$-norm in terms of the $a$-weighted Sobolev-type norm of the form $\|g\|_\Gamma \leq \|g\|_{\nu,a}$.*

*Remark.* By scaling one may construct examples that show one can not do with less than the first $\nu$ derivatives, in general.

*Proof.* Put disjoint balls of radius $a/2$ around the points of $\Gamma$. Consider anyone of them. And assume without loss that its center is $z = 0$. Write for simplicity $B \equiv B_{a/2}(0)$. Then we have $|g(0)| \times |B| \leq \int_B |g(y) - g(0)| dy + \int_B |g(y)| dy$. To express the first integral on the r.h.s. as an integral of the derivatives of $g$ over $B$ we use Polar coordinates $\int \Omega(de) \int_0^{a/2} dr\, r^{\nu-1} |g(re) - g(0)|$. We use the one-dimensional Taylor-expansion of the function $r \mapsto g(re) =: \chi_e(r)$ of the radial coordinate $r$ up to order $\nu - 1$. Expanding around the point $r$ we get $|g(re) - g(0)| \leq \sum_{k=1}^{\nu-1} \frac{r^k}{k!} |\chi_e^{(k)}(r)| + \int_0^r ds \frac{s^{\nu-1}}{(\nu-1)!} |\chi_e^{(\nu)}(s)|$. This gives $\int_B |g(y) - g(0)| dy \leq \sum_{k=0}^{\nu} \frac{(a/2)^k}{k!} \int \Omega(de) \int_0^{a/2} dr\, r^{\nu-1} |\chi_e^{(k)}(r)|$. The reader should check that also the term for $k = \nu$ can be bounded in this form (interchange the orders of integration between $s$ and $r$!) This argument only works since the power $\nu - 1$ reappears under the integral of the remainder term. Dividing this inequality by the volume of $B$, bounding the $k$-th directional derivatives by $\|d^k g\|$, and integrating over the whole of $\mathbb{R}^\nu$ now proves the claim.

For the semi-norm $\|g\|_{\Gamma,\delta}$ that was introduced in (3.1) (needed to control Model B) we get the following analogous estimate.

**Proposition 4.** *Suppose that $\Gamma \subset \mathbb{R}^\nu$ and the number $a$ is a bound on the minimal distance between the points in $\Gamma$. Assume that $\tilde{a} := a - 4\delta > 0$. Then we have the bound in terms of the $\tilde{a}$-weighted Sobolev-type semi-norm*

$$\|g\|_{\Gamma,\delta} \leq 4\delta \frac{1}{|B_1|} \sum_{k=0}^{\nu} \frac{1}{k!} \frac{1}{(\tilde{a}/2)^{\nu-k}} \int_{\mathbb{R}^\nu \setminus B_{a/2}(0)} \|d^{k+1} g(y)\| dy \quad \left( \leq 4\delta \|dg\|_{\nu,\tilde{a}} \right).$$

*Proof.* For fixed $x \neq y$ in $\Gamma$ and any $|z|, |z'| \leq 2\delta$ we have $|g(x - y + z) - g(x - y + z')| \leq \int_0^{|z'-z|} \sup_{|e|=1} \left|\frac{d}{dt} g(x - y + z + te)\right| dt \leq 4\delta \sup_{w \in B_{2\delta}(x-y)} \sup_{|e|=1} \left|\frac{d}{dt}\right|_{t=0} g(w + te)\right|$.

Using the estimate in terms of the integrals over balls in terms of derivatives up to order of the dimension provided in the proof of Proposition 3 we get for $w \in B_{2\delta}(x - y)$
$\left|\frac{d}{dt}\right|_{t=0} g(w + te)\right| \leq \frac{1}{|B_1|} \sum_{k=0}^{\nu} \frac{1}{k!} \frac{1}{(\tilde{a}/2)^{\nu-k}} \int_{B_{\tilde{a}/2}(w)} \|d^k \frac{d}{dt}\|_{t=0} g(u + te)\| du$

Here we have used the radius $\tilde{a}/2$ because this implies that $B_{\tilde{a}/2}(w) \subset B_{a/2}(x - y)$, independently of $w$, and so we get that the r.h.s is bounded by $\frac{1}{|B_1|} \sum_{k=0}^{\nu} \frac{1}{k!} \frac{1}{(\tilde{a}/2)^{\nu-k}} \int_{B_{a/2}(x-y)} \|d^{k+1} g(u)\| du$. This gives the desired estimate by summing over $y$ that are not equal to $x$, and extending the integral over all of $\mathbb{R}^\nu \setminus B_{a/2}(0)$.

**Acknowledgments:** I am grateful to M.Baake for intriguing me with random diffraction patterns and pointing out references [BaaHoe00], [Hof95b]. This work was supported by the DFG Schwerpunkt 'Wechselwirkende stochastische Systeme hoher Komplexität'.